\documentclass[twoside,11pt]{article}

\usepackage[cp1251]{inputenc}
\usepackage{graphicx}

\usepackage{amssymb}
\usepackage{amsmath}
\usepackage{amsfonts}
\usepackage{wasysym}

\begin{document}           

\title{\Large Dark energy and light WIMP dark matter from the cosmological viewpoint}
\author{V.E. Kuzmichev, V.V. Kuzmichev\\[0.5cm]
\itshape Bogolyubov Institute for Theoretical Physics,\\
\itshape National Academy of Sciences of Ukraine, Kiev, 03680 Ukraine}

\date{}

\maketitle

\begin{abstract}
We give an estimation of the masses of light dark matter particle and dark energy quasiparticle which can be extracted from the astrophysical data about the contributions of baryon ($\Omega_{B}$), dark matter ($\Omega_{DM}$), and dark energy ($\Omega_{X}$) densities to the total matter-energy density budget in our universe. We use the quantum cosmological model in which dark energy is a condensate of quantum oscillations of
primordial scalar field. The dark energy quasiparticle with the mass $\approx 15$ GeV is consistent with  the 7-year WMAP and other data on $\Omega_{B}$ and 
$\Omega_{X}$. The quasiparticles can decay with violation of CP-invariance into baryons, leptons, and dark matter. The WIMP mass $\approx 5$ GeV corresponds to observed values of 
$\Omega_{B}$ and $\Omega_{DM}$. Such a mass agrees with the observations of CoGeNT, DAMA, and CDMS.
Quasiparticles of dark energy can be registered as a constant background of radiation with the frequency $\approx 3.64 \times 10^{24}$ s$^{-1}$. Dark matter particles must exhibit themselves in the form of signals with the frequency  $\approx 1.21 \times 10^{24}$ s$^{-1}$ of  radiation from galaxy clusters where the intensive gravitational fields produced by dark matter
exist.

\end{abstract}


\ \\ [1cm]

\textbf{1.} The CoGeNT collaboration \cite{CoG} has recently announced the observation of an excess of events at low energies which can be interpreted as a detection of dark matter particles with a mass in the range of $m_{\chi} \sim 7 - 11$ GeV. Account of the observations of DAMA collaboration \cite{DAMA} and the CDMS experiment \cite{CDMS} allows to establish the restriction on a mass of dark matter particle in the range of $m_{\chi} \sim 5 - 10$ GeV or even $m_{\chi} \sim 1 - 10$ GeV within the context of asymmetric dark matter models (ADM) \cite{FHZ}. Assuming the value $\sigma \approx 10^{-40}$ cm$^{-2}$ for an elastic scattering cross section of dark matter particle with nucleus and using the measured dark matter abundance $\Omega_{DM} h^{2} = 0.11$ \cite{WMAP7}, with the help of the effectively coupling analysis one can conclude that the observed signals should be interpreted as an evidence for a scalar WIMP with scalar interaction.

The low mass dark matter problem can be analyzed, e.g., within the context of the standard model with scalar dark matter, ADM models, or the minimal supersymmetric standard model with neutralino dark matter (see, for instance, bibliographies in Refs \cite{CoG,FHZ} and Refs \cite{AA,FZN,BDFS}).
A different model based on the quantum cosmological approach involving the available data on the abundance of baryons and dark matter in our universe was proposed in Ref. \cite{K1}. Obtained restriction on the mass of dark matter particle with zero spin $m_{\chi} < 15$ GeV, with the preference for $m_{\chi} \sim 5 - 10$ GeV, agrees with the values given in Refs \cite{CoG,FHZ}. 

\textbf{2.} In this paper we analyze the possible properties of dark energy and dark matter particles (such as mass, spin, and charges) using the  WMAP7-year and other data \cite{WMAP7}. We consider the nonstationary universe in which dark energy is a source of observed matter in the form of baryons, leptons, and dark matter, while dark energy itself is identified with a condensate of quantum oscillations of primordial scalar field $\phi$ filling the homogeneous, isotropic, and closed universe. 
According to the quantum cosmological model \cite{K5} this condensate is an antigravitating quantum medium of excitation $\phi$-quanta (zero momentum quasiparticles) of the spatially coherent oscillations of primordial scalar field about an equilibrium state $\sigma$ corresponding to (true, $\Lambda =0$, or false, $\Lambda \neq 0$) vacuum.
Its energy density can be written as follows
\begin{equation}\label{1}
    \rho_{k} = \frac{2M_{k}}{a^{3}}, \qquad 
     M_{k} = m_{\phi}\left(k + \frac{1}{2}\right),
\end{equation}
where $a$ is the cosmic scale factor, $M_{k}$ is an amount of matter (mass) in the universe related to quantized scalar field, $k$ is the number of $\phi$-quanta with the mass $m_{\phi} = \left[\frac{d^{2}V(\phi)}{d\phi ^{2}}\right]_{\sigma}^{1/2}$, $V(\phi)$ is the potential of the scalar field.

Here we use the modified Planck system of units in which $l_{P} = \sqrt{\frac{2G\hbar}{3\pi c^{3}}}$ is taken as a unit of length, $m_{P} = \frac{\hbar}{l_{P} c}$ is a unit of mass, $\rho_{P} = \frac{3c^{4}}{8\pi G l_{P}^{2}}$ is a unit of energy density and so on. In this units a comoving volume is equal to $\frac{1}{2}\,a^{3}$, and the dimensionless critical density equals to the dimensionless squared Hubble expansion rate $H^{2}$. 

The energy density of a condensate relative to critical density is
\begin{equation}\label{2}
    \Omega_{\phi} = \frac{2M_{k}}{a^{3}H^{2}}.
\end{equation} 

According to the model under consideration the matter energy density $\Omega_{\phi}$ on large spacetime scales has to be represented by the sum of the terms
\begin{equation}\label{4}
    \Omega_{\phi} = \Omega_{B} + \Omega_{L} + \Omega_{DM} + \Omega_{CMB} + \Omega_{X},
\end{equation}
where $\Omega_{B}$, $\Omega_{L}$, $\Omega_{DM}$, and $\Omega_{CMB}$ are the energy density of barions, leptons, dark matter particles, and the cosmic microwave background radiation (CMB), $\Omega_{X}$ is the density of residual dark energy (i.e. portion of a condensate which has not decayed to the moment of observation, for details see below). All quantities in Eqs (\ref{2}) and (\ref{4}) relate to some fixed instant of time $t$ which is an argument of $a$ and $H$.

Eqs (\ref{2}) and (\ref{4}) can be made consistent with each other if one assumes that under the action of gravitational force the $\phi$-quanta decay via a two-step process into particles
\begin{eqnarray}\label{5}
    \phi \rightarrow \chi + \nu + n &  & \nonumber \\
                                    & \searrow & \nonumber \\
                                    &  & p + e^{-} + \bar{\nu},
\end{eqnarray}
or antiparticles
\begin{eqnarray}\label{6}
  \phi = \bar{\phi} \rightarrow \bar{\chi} + \bar{\nu} + \bar{n} 
                                                           & & \nonumber \\
                                                           & \searrow & \nonumber \\
                                                           &  & \bar{p} + e^{+} + \nu,  
\end{eqnarray}
where we have used the standard notations for neutron, proton, electron, positron, and neutrino, while the bars denote antiparticles. The $\chi$ is the quantum of the residual excitation of the oscillations of the field $\phi$ about equilibrium state. The set of $\chi$-particles forms dark matter. Neutrino takes away the spin and the dark matter particle $\chi$ may have spin 0 or 1. It follows from Ref. \cite{FHZ} that dark matter particle may be a scalar particle with scalar interaction. Therefore one can accept that the spin of $\chi$ is equal to zero. Since the universe  is a charge neutral system, then all charges of the $\phi$-quantum may be taken equal to zero. As it is assumed that dark matter has a non-baryonic nature, then the baryonic charge of $\chi$ is equal to zero. According to the reactions (\ref{5}) and (\ref{6}), the $\phi$-quantum decays into three particles or antiparticles with the subsequent decay of $n$ or $\bar{n}$ into three more particles. The inverse fusion reaction of three or more particles into the zero-momentum $\phi$-quantum is negligibly small. As a result the T-invariance breaks down in the decays (\ref{5}) and (\ref{6}), and the arrow of time arises.

Let us suppose that this theory is CPT-invariant. Then the violation of T-invariance leads to the violation of CP-invariance as well. According to modern view, CP-violation can be caused by the baryon asymmetry of the universe. 

Particles and antiparticles of the decays (\ref{5}) and (\ref{6}) can annihilate between themselves and contribute to the observed CMB. In the decays (\ref{5}) and (\ref{6}) five different ``particle-antiparticle'' pairs are produced. Since $\nu\bar{\nu}$ scattering cross section is significantly smaller than  $p\bar{p}$ scattering cross section, then on the output we shall have $n_{\gamma} = 6$ photons and $n_{\nu} = 4$ neutrinos/antineutrinos. The ratio $\frac{n_{\nu}}{n_{\gamma}} \approx 0.67$ can be compared with the observed value $\frac{n_{\nu}}{n_{\gamma}}|_{today} \approx 0.9$. Then it follows that $\sim 26\%$ of the registered neutrinos have secondary origin (e.g. may come from stars). In order to calculate the parameter $\eta = \frac{n_{B}}{n_{\gamma}} \sim 10^{-10}$, where $n_{B}$ is number of baryons, the quantum model of a condensate with CP-violation has to be constructed. This question is outside of the scope of this paper.

\textbf{3.} Let us consider the decay channels (\ref{5}) and (\ref{6}) with regard for baryon asymmetry of the universe. Since $\Omega_{L} \ll \Omega_{B}$, then leptons may be neglected, neutrino and antineutrino may be included in dark matter contribution term $\Omega_{DM}$, and annihilation photons in the density $\Omega_{CDM}$. We suppose that the $\phi$-quanta and neutrons decay independently with some decay rates $\Gamma_{\phi}$ and $\Gamma_{n} = \Gamma_{\bar{n}}$. According to the quantum model description \cite{K5}, when the universe expands, the following condition is realized at every instant of time for large enough number of the $\phi$-quanta,
\begin{equation}\label{7}
    \langle a \rangle_{k} = M_{k}, 
\end{equation}  
where $\langle a \rangle_{k}$ is the mean value of the scale factor $a$ in the k-state of the universe with the mass of a condensate $M_{k}$. Eq. (\ref{7}) can be interpreted as a mathematical formulation of the Mach's principle proposed by Sciama \cite{Di,K8}.  In the classical limit the evolution of the mean value $\langle a \rangle_{k}$ in time is described by the Einstein-Friedmann equations. Denoting the number of the $\phi$-quanta as a function of the proper time $t$ by $N_{\phi}(t)$ and taking into account Eq. (\ref{1})  we can write the condition (\ref{7}) as follows
\begin{equation}\label{8}
    a(t) = m_{\phi} N_{\phi}(t).
\end{equation}
Under expansion of the universe in accordance with the Hubble law
\begin{equation}\label{9}
    \frac{da(t)}{dt} = H(t) a(t),
\end{equation}
the number of the $\phi$-quanta will change as
\begin{equation}\label{10}
    \frac{dN_{\phi}(t)}{dt} = H(t) N_{\phi}(t).
\end{equation}
Then taking into account two-channel $\phi$-quantum decay as in the schemes (\ref{5}) and (\ref{6}), we can write the balance equations
\begin{eqnarray}\label{11} 
    \frac{d\delta N_{\phi}(t)}{dt} = -\,\lambda (t)\delta N_{\phi}(t), \quad
     \frac{d\delta N_{n}(t)}{dt} = -\,\Gamma _{n}(t)\delta N_{n}(t) + 
                                 \Gamma _{\phi}(t)\delta N_{\phi}(t), \nonumber \\
     \frac{d\delta N_{p}(t)}{dt} = -\,\Gamma _{p}(t)\delta N_{p}(t) +
        \Gamma _{n}(t)\delta N_{n}(t),
\end{eqnarray}
where $d\delta N_{\phi} = dN_{\phi} - dN_{\bar{\phi}}$  is the difference between the $\phi$-quanta decaying into particles and antiparticles in the channels (\ref{5}) and (\ref{6}) during the time $dt$, 
\begin{equation}\label{12}
    \lambda(t) = \Gamma_{\phi}(t) - H(t)
\end{equation}
is an effective decay rate which takes into account the change of the number of the $\phi$-quanta during the expansion of the universe, $\delta N_{n} = N_{n} -  N_{\bar{n}}$ and $\delta N_{p} = N_{p} -  N_{\bar{p}}$ are net amounts of neutrons and protons respectively at some instant of time $t$. Taking $\Gamma _{p} = \Gamma _{\bar{p}} = 0$ and choosing the initial conditions as
\begin{equation}\label{13}
    \delta N_{\phi}(t') = N, \quad \delta N_{n}(t') = 0, \quad \delta N_{p}(t') = 0,
\end{equation}
where $N$ is the number of the $\phi$-quanta at some arbitrary chosen initial instant of time $t'$, we find the solution of the set (\ref{11})
\begin{equation}\label{14}
    \frac{\delta N_{\phi}(t)}{N} = e^{- \overline{\lambda}\Delta t},
\end{equation}
\begin{equation}\label{15}
    \frac{\delta N_{n}(t)}{N} = \int_{t'}^{t}\!\!dt_{1} \Gamma_{\phi}(t_{1})\,
e^{- \int_{t'}^{t_{1}}\!\!dt_{2} \lambda(t_{2})}\,
e^{- \int_{t_{1}}^{t}\!\!dt_{2} \Gamma_{n}(t_{2})},
\end{equation}
\begin{equation}\label{16}
    \frac{ \delta N_{p}(t)}{N} = \int_{t'}^{t}\!\!dt_{1} \Gamma_{n}(t_{1})
\int_{t'}^{t_{1}}\!\!dt_{2} \Gamma_{\phi}(t_{2})\,
e^{- \int_{t'}^{t_{2}}\!\!dt_{3} \lambda(t_{3})}\,
e^{- \int_{t_{2}}^{t_{1}}\!\!dt_{3} \Gamma_{n}(t_{3})},
\end{equation} 
where $\overline{\lambda} = \overline{\Gamma}_{\phi} - \overline{H}$ and
\begin{equation}\label{17}
    \overline{\Gamma}_{\phi} = \frac{1}{\Delta t}
\int_{t'}^{t}\!\!dt_{1} \Gamma_{\phi}(t_{1}), \qquad
     \overline{H} = \frac{1}{\Delta t} \int_{t'}^{t}\!\!dt_{1} H(t_{1})
\end{equation}
are the mean decay rate of the $\phi$-quantum and the mean Hubble expansion rate on the time interval $\Delta t = t - t'$.

The decay rates of the $\phi$-quanta $\Gamma_{\phi}(t)$ and neutrons $\Gamma _{n}(t)$ are unknown. We shall assume that these decay rates depend very weakly on averaging interval. Then from Eqs (\ref{15}) and (\ref{16}) it follows the simple expressions
\begin{equation}\label{18}
     \frac{\delta N_{n}(t)}{N} = 
\frac{\overline{\Gamma}_{\phi}}{\overline{\Gamma}_{n} - \overline{\lambda}}\,
\left(e^{-\overline{\lambda}\Delta t} - e^{-\overline{\Gamma}_{n}\Delta t}\right),
\end{equation} 
\begin{equation}\label{19}
     \frac{ \delta N_{p}(t)}{N} = \frac{\overline{\Gamma}_{\phi}}{\overline{\lambda}}\,
\left[1 + \frac{1}{\overline{\Gamma}_{n} - \overline{\lambda}}\, 
\left(\overline{\lambda}\,e^{-\overline{\Gamma}_{n}\Delta t} -  
\overline{\Gamma}_{n}\,e^{-\overline{\lambda}\Delta t}  \right) \right].
\end{equation}
For our universe today $\overline{\Gamma}_{n} = 1.12 \times 10^{-3}$ s$^{-1}$, $H_{0} = 71.0 \pm 2.5 \,\mbox{km}\, \mbox{s}^{-1}\mbox{Mpc}^{-1}$ and the age $t_{0} = 13.75 \pm 0.13$ Gyr \cite{WMAP7}. Taking $\Delta t = t_{0}$ for estimation we find that
\begin{equation}\label{20}
    H_{0}\Delta t = 0.999, \qquad \overline{\Gamma}_{n}\Delta t = 4.86 \times 10^{14}. 
\end{equation}
We have supposed that the decay of the $\phi$-quanta is caused mainly by the action of gravitational forces. Then the inequality $\overline{\Gamma}_{\phi} \ll \overline{\Gamma}_{n}$ must hold, and we can put
\begin{equation}\label{21}
    \overline{\lambda} \ll \overline{\Gamma}_{n}.
\end{equation}
Under this condition the number of baryons $\delta N_{p}$ in the expanding universe obeys the law
\begin{equation}\label{22}
    \frac{\delta N_{p}(t)}{N} = \frac{\overline{\Gamma}_{\phi}}{\overline{\lambda}}\,
     \left[1 - e^{-\overline{\lambda}\Delta t} \right].   
\end{equation}

\textbf{4.} The baryon density at the instant of time $t$ is equal to
\begin{equation}\label{23}
    \Omega_{B} = \frac{2m_{p}\delta N_{p}(t)}{a^{3}(t) H^{2}(t)},
\end{equation}
where $m_{p}$ is the proton mass. The density of dark energy which has not decayed at the instant of time $t$ is 
\begin{equation}\label{24}
     \Omega_{X} = \frac{2m_{\phi}\delta N_{\phi}(t)}{a^{3}(t) H^{2}(t)}.
\end{equation}
The ratio $\frac{\Omega_{B}}{\Omega_{X}}$ can be written as follows
\begin{equation}\label{25}
    \frac{\Omega_{B}}{\Omega_{X}} = \sqrt{\frac{g_{p}}{g_{\phi}}}\,\,
\frac{\overline{\Gamma}_{\phi}}{\overline{\lambda}}\,
 \left[e^{\overline{\lambda}\Delta t } - 1\right],
\end{equation}
where $g_{p} = G m_{p}^{2}$ and $g_{\phi} = G m_{\phi}^{2}$ are the dimensionless gravitational coupling constants for proton and $\phi$-quantum. We note that the  coupling constant $g_{p} = 0.59 \times 10^{-38}$ is very small. The decay rate $\overline{\Gamma}_{\phi}$ has to be proportional to the coupling constant $g_{\phi}$ and the dark energy number density $|\psi (0)|^{2}$, where $\psi$ is the wave  function of the $\phi$-quantum \cite{K1}. So the expression (\ref{25}) can be rewritten in the form
\begin{equation}\label{26}
     \frac{\Omega_{B}}{\Omega_{X}} = \sqrt{\frac{g_{p}}{g_{\phi}}}\,\,
\frac{g_{\phi} \gamma}{g_{\phi} \gamma - \beta}\,
 \left[e^{g_{\phi} \gamma - \beta } - 1\right],
\end{equation}
where we denote $\beta \equiv \overline{H}\Delta t$ and $g_{\phi} \gamma \equiv \overline{\Gamma}_{\phi} \Delta t$. The ratio (\ref{26}) as the function of $g_{\phi}$ has the properties
\begin{equation}\label{27}
    \frac{\Omega_{B}}{\Omega_{X}}\left|_{g_{\phi} \sim 0}\right. \sim \sqrt{g_{p} g_{\phi}}, \qquad \frac{\Omega_{B}}{\Omega_{X}}\left|_{g_{\phi} \rightarrow \infty}\right. \sim 
\sqrt{\frac{g_{p}}{g_{\phi}}}\,\, e^{g_{\phi} \gamma},
\end{equation}
and 
\begin{equation}\label{28}
    \frac{\Omega_{B}}{\Omega_{X}}\left|_{g_{\phi} \gamma = \beta}\right. = 
\sqrt{\frac{g_{p}}{g_{\phi}}}\,\,\overline{H} \Delta t.
\end{equation}
The ratio (\ref{28}) describes the case $\overline{\Gamma}_{\phi} \Delta t = \overline{H} \Delta t$  when  the number of the $\phi$-quanta is constant, $\delta N_{\phi}(t) = N = \mbox{const.}$ The coupling constant $g_{\phi}$ is equal to
\begin{equation}\label{29}
    g_{\phi} = g_{p} \,\left(\frac{\Omega_{X}}{\Omega_{B} }\,
\overline{H}\Delta t\right)^{2}_{\overline{\lambda}=0}.
\end{equation}

Using the WMAP and other data \cite{WMAP7}, $\Omega_{B} = 0.0456 \pm 0.0016$, $\Omega_{X} = 0.728^{+0.015}_{-0.016}$, and Eq. (\ref{20}), and setting $\overline{H} \Delta t = 
H_{0} \Delta t$,  we obtain
\begin{equation}\label{30}
    g_{\phi} \approx 254\, g_{p}.
\end{equation}
Under this assumption the decay rate of the $\phi$-quantum is equal to $\overline{\Gamma}_{\phi} = 2.30 \times 10 ^{-18}$ s$^{-1}$. It is close to the value found in Ref. \cite{K1} in a model which does not take into account Eq. (\ref{10}), thus being equivalent to the case $\overline{\Gamma}_{\phi} \gg \overline{H}$. In the opposite case, when $\overline{\Gamma}_{\phi} \ll \overline{H}$, the ordinary matter will not be produced in the universe. Therefore the value (\ref{30}) may be considered as realistic.

The coupling constant (\ref{30}) corresponds to the mass
\begin{equation}\label{31}
    m_{\phi} \approx 16\, m_{p}, \qquad \mbox{or} \qquad 
 m_{\phi} \approx 15\,\mbox{GeV}.
\end{equation}
The quasiparticles with such a mass, if they exist, have to be observed as a constant component of radiation with the frequency 
\begin{equation}\label{32}
    \nu_{\phi} \approx 3.64 \times 10^{24}\,\mbox{s}^{-1},
\end{equation}
which is the same in every direction from the registration place. 

\textbf{5.} Energy density of dark matter at the instant of time $t$ is
\begin{equation}\label{33}
    \Omega_{DM} = \frac{2m_{\chi }\delta N_{\chi}(t)}{a^{3}(t) H^{2}(t)}.
\end{equation}
Taking into account that the number of dark matter particles $\chi$ according to the decay schema (\ref{5}) is equal to the number of baryons,
\begin{equation}\label{34}
    \delta N_{\chi}(t) = \delta N_{p}(t),
\end{equation}
we find that
\begin{equation}\label{35}
    \frac{\Omega_{B}}{m_{p}} = \frac{\Omega_{DM}}{m_{\chi}}.
\end{equation}

For the observed values of $\Omega_{B}$ and $\Omega_{DM} = 0.227 \pm 0.014$ \cite{WMAP7} it follows that
\begin{equation}\label{37}
    \frac{\Omega_{DM}}{\Omega_{B}} \approx 5, \qquad \mbox{or}
\qquad m_{\chi} \approx 5\, m_{p}.
\end{equation}
With regard to the errors of measurement of $\Omega_{DM}$ and $\Omega_{B}$ one can take the value of the mass of the $\chi$-particle equal to
\begin{equation}\label{38}
    m_{\chi} \approx 5\,\mbox{GeV}.
\end{equation}
This value is in the range $m_{\chi} \sim 1 - 10$ GeV indicated in Ref. \cite{FHZ}. 
It agrees with the observations of CoGeNT \cite{CoG}, DAMA \cite{DAMA}, and CDMS \cite{CDMS}.
Eq. (\ref{34}) is a relationship between the dark matter and baryon chemical potentials with precise value $c_{1} = 1$ of the coefficient $c_{1}$ introduced in ADM models (cf. \cite{FHZ}).

The following condition between dark matter density and the mass of particles which form dark matter has to be fulfilled 
\begin{equation}\label{39}
    \Omega_{DM} = (0.0486 \pm 0.017) \,m_{\chi},
\end{equation}
where $m_{\chi}$ is in GeV.
In Fig.~1 the region of admissible values of $m_{\chi}$ for the observed scattered values of $\Omega_{DM}$ is shown. The slope of a straight line (\ref{39}) determines the density $\Omega_{B}$.

\begin{figure}[ht]
\begin{center}
\includegraphics*[width=0.9\textwidth]{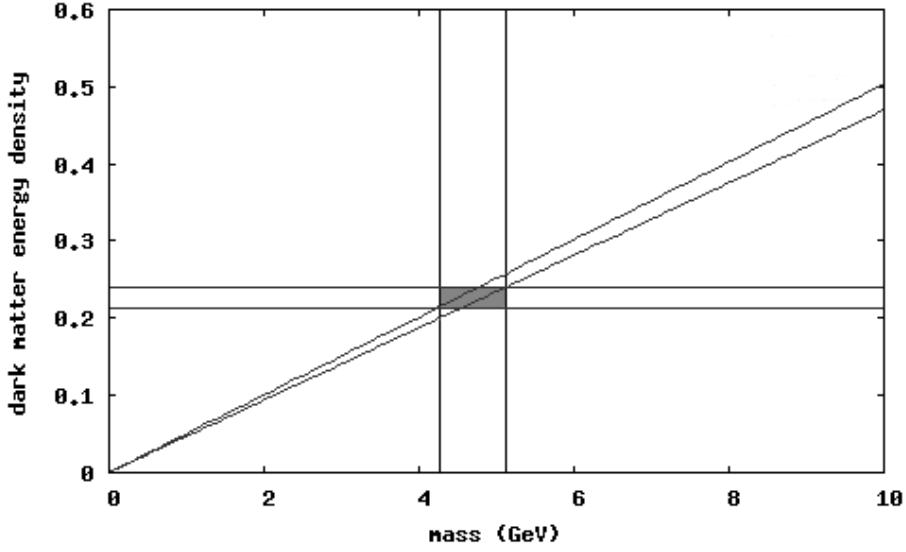}
\end{center}
\caption{The energy density $\Omega_{DM}$ \textit{versus} the dark matter particle mass $m_{\chi}$ in GeV. The observed density $\Omega_{DM} = 0.227 \pm 0.014$ \cite{WMAP7} corresponds to the mass $m_{\chi} = 4.69 \pm 0.45$ GeV.} \label{fig:1}
\end{figure}

The particle with the mass (\ref{38}) is characterized by the frequency in the energy spectrum $\nu_{DM}$ and the length of the  de Broglie's wave $\lambda_{DM}$ equal to
\begin{equation}\label{40}
   \nu_{DM} =\left[1 + \frac{1}{2} \left(\frac{v_{DM}}{c}\right)^{2}\right] \, 1.21 \times 
              10^{24}\,\mbox{s}^{-1}, \quad
    \lambda_{DM} = \left(\frac{c}{v_{DM}}\right)\,2.48 \times 10^{-14}\, \mbox{cm},
\end{equation}
where $v_{DM}$ is  the velocity of dark matter particle, $c$ is the light velocity, $v_{DM} < c$. 

The values of masses (\ref{31}) and (\ref{38}) show that the decay (\ref{5}) occurs with release of energy $\approx 9$ GeV in the form of kinetic energy of decay products.

From Eqs (\ref{24}) and (\ref{33}) it follows 
\begin{equation}\label{41}
    \frac{\Omega_{X}}{\Omega_{DM}} = \frac{m_{\phi} \delta N_{\phi}}{m_{\chi} \delta N_{\chi}}.
\end{equation}
Using the observed values of $\Omega_{X}$ and $\Omega_{DM}$, and the obtained values of the masses $m_{\phi}$ (\ref{31}) and $m_{\chi}$ (\ref{37}), we find that
\begin{equation}\label{42}
    \delta N_{\phi} \approx \delta N_{\chi}.
\end{equation}
It means that the chemical potentials of dark energy and dark matter coincide. Eq. (\ref{42}) is a manifestation of so-called coincidence problem ($\frac{\Omega_{X}}{\Omega_{DM}} \approx 3.2$). Its solution is a separate non-trivial task (see, e.g., Ref. \cite{LS} and references therein).

\textbf{6.} Thus one can conclude that the observed values of the densities $\Omega_{B}$, $\Omega_{DM}$, and $\Omega_{X}$ in the model of the decays (\ref{5}) and (\ref{6}) lead to the values of mass $\sim 15$ GeV of dark energy quasiparticle and mass $\sim 5$ GeV of dark matter particle. Both these masses can be observed, the first one in the form of constant background of radiation with the frequency close to (\ref{32}), while the second one in the form of signals with the frequency and wave length of radiation close to (\ref{40}) from galaxy clusters where the intensive gravitational fields produced by dark matter exist.


\begin{thebibliography}{99}
\itemsep -6pt plus 1pt minus 1pt
\bibitem{CoG}C.E. Aalseth et al. (CoGeNT), arXiv:1002.4703 [astro-ph.CO] (2010).

\bibitem{DAMA}R. Bernabei et al. (DAMA), Eur. Phys. J. C \textbf{56}, 333 (2008).

\bibitem{CDMS}Z. Ahmed et al. (CDMS), arXiv:0912.3592 [astro-ph.CO] (2009).

\bibitem{FHZ}A.L. Fitzpatrick, D. Hooper, K.M. Zurek, arXiv:1003.0014 [hep-ph] (2010).

\bibitem{WMAP7}N. Jarosik et al. (WMAP7), arXiv:1001.4744 [astro-ph] (2010).

\bibitem{AA}S.Andreas et al, arXiv:1003.2595 [hep-ph] (2010).

\bibitem{FZN} D. Feldman, Z. Liu, P. Nath, arXiv:1003.0437 [hep-ph] (2010).

\bibitem{BDFS} A. Bottino, F. Donato, N. Fornengo, S. Scopel, arXiv:0912.4025 [hep-ph] (2009); Phys. Rev. D \textbf{78}, 083520 (2008).

\bibitem{K1}V.E. Kuzmichev, V.V. Kuzmichev, in: Trends in Dark Matter 
Research, ed. J.V. Blain, Nova Science, Hauppage 2005 [astro-ph/0405455];
Ukr. J. Phys \textbf{48}, 801 (2003) [arXiv:astro-ph/0301017]]; 
arXiv:astro-ph/0302173 (2003); in: Selected Topics in Theoretical Physics and Astrophysics, eds. A.K. Motovilov, F.M.Pen'kov, JINR, Dubna 2003.

\bibitem{K5}V.E. Kuzmichev, V.V. Kuzmichev, Acta Phys. Pol. B
\textbf{39}, 2003 (2008) [arXiv:0712.0465 [gr-qc]];
ibid.
\textbf{39}, 979 (2008) [arXiv:0712.0464 [gr-qc]];
ibid.
\textbf{40}, 2877 (2009) [arXiv:0905.4142 [gr-qc]].

\bibitem{Di}R.H.Dicke, in: Gravitation and Relativity, eds. Hong-Yee Chiu, W.F. Hoffmann, Benjamin, New York 1964.

\bibitem{K8}V.E. Kuzmichev, V.V. Kuzmichev, Ukr. Phys. J. \textbf{50}, 1321 (2005) [arXiv:astro-ph/0510763].

\bibitem{LS} G. Leon, E. N. Saridakis, arXiv:0904.1577 [gr-qc] (2009).

\end{thebibliography}
\end{document}